# Packet Header Recognition Utilizing an All-Optical Reservoir Based on Reinforcement-Learning-Optimized Double-Ring Resonators

Zheng Li, Xiaoyan Zhou, Zongze Li, Guanju Peng, Yuhao Guo, and Lin Zhang

*Abstract*—Optical packet header recognition is an important signal processing task of optical communication networks. In this work, we propose an all-optical reservoir, consisting of integrated double-ring resonators (DRRs) as nodes, for fast and accurate optical packet header recognition. As the delay-bandwidth product (DBP) of the node is a key figure-of-merit in the reservoir, we adopt a deep reinforcement learning algorithm to maximize the DBPs for various types of DRRs, which has the advantage of full parameter space optimization and fast convergence speed. Intriguingly, the optimized DBPs of the DRRs in cascaded, parallel, and embedded configurations reach the same maximum value, which is believed to be the global maximum. Finally, 3-bit and 6-bit packet header recognition tasks are performed with the all-optical reservoir consisting of the optimized cascaded rings, which have greatly reduced chip size and the desired "flat-top" delay spectra. Using this optical computing scheme, word-error rates as low as $5 \times 10^{-4}$ and $9 \times 10^{-4}$ are achieved for 3-bit and 6-bit packet header recognition tasks, respectively, which are one order of magnitude better than the previously reported values.

## I. INTRODUCTION

THE rapid growth of data in optical communications is promoting a change in the network architecture from optical circuit switching to optical packet switching (OPS), which allows a higher bandwidth with a lower power consumption [1], [2]. An optical packet consists of an optical header with routing information and a payload [3], [4]. Thus, optical packet header recognition becomes essential in OPS networks, which determines whether an OPS router sends the

Manuscript received XXX; revised XXX; accepted XXX. Date of publication XXX; date of current version XXX. This work was supported by National Natural Science Foundation of China under grant 62005195 and 62201305. (*Corresponding authors: Xiaoyan Zhou; Lin Zhang.*)

Zheng Li, Xiaoyan Zhou, Guanju Peng, and Lin Zhang are with the Tianjin Key Laboratory of Integrated Opto-electronics Technologies and Devices, School of Precision Instruments and Opto-electronics Engineering, Tianjin University, Tianjin 300072, China (e-mail: zhemglee@tju.edu.cn; xiaoyan_zhou@tju.edu.cn; guanjup@tju.edu.cn; lin_zhang@tju.edu.cn).

Xiaoyan Zhou, Zongze Li, and Lin Zhang are with the Peng Cheng Laboratory, Shenzhen 518038, China (e-mail: xiaoyan_zhou@tju.edu.cn; lizz@pcl.ac.cn; lin_zhang@tju.edu.cn).

Yuhao Guo is with the Huawei Technologies Co., Ltd., Shenzhen 518129, China (e-mail: guoyuhao2@huawei.com).

Color versions of one or more of the figures in this article are available online at http://ieeexplore.ieee.org

payload to the correct output port. Current techniques for packet header recognition mainly rely on optoelectronic devices, in which optical-electrical-optical (OEO) conversion inevitably constrains the processing speed of a recognition system [5], [6]. Recently, all-optical packet head recognition systems based on reservoir computing have been reported, circumventing the OEO conversion [7]-[14], in which optical delays are key components and have been realized by bulk optics [7], active devices [8]-[10], fibers [11], virtual neurons [12], and waveguides [13], [14]. These devices typically have a large footprint or a high power-consumption. In contrast, microring resonators as a delay element [15]-[19] would be promising for power-efficient on-chip optical packet header recognition.

Although a delay element is often critically important for signal processing [20], [21], including packet header recognition, one should note that the delay-bandwidth product (DBP) of a delay element may be a limiting factor of its performance, especially when output signal quality is a key consideration [22], [23]. The implementation of microring resonators with a large group delay ($\tau$) and a broad bandwidth ($\Delta\omega$), i.e., a large DBP, plays a key role [24]. Typically, the resonance enhancement of group delay in a single resonator comes at the expense of reduced bandwidth, that is, the DBP of a single resonator is restricted by a constant ($C$), i.e., $DBP = \tau \times \Delta\omega \leq C$ [24]. In principle, using multiple resonators, one may obtain a larger value of $C$ in the DBP [25], [26]. Nevertheless, the upper limit of a multi-ring system's DBP remains unexplored.

Genetic algorithms [24] may be used to optimize a single microring for a large DBP, often with heavy computation loads or inevitable convergence errors [24], partially due to the continuous-variable feature of ring resonators' parameter space. Recently, artificial intelligence (AI) technique has emerged as a promising approach for resource allocation in optical networks [27], [28] and for designing optical functional units [29], [30], offering fast convergence speed and low convergence errors. In particular, deep reinforcement learning (DRL) provides a framework that learns to solve complex problems through a trial-and-error process [31], which has been proven highly scalable for various model-free problems [29]-[33], with applications ranging from optimization of the dielectric nanostructures in a solar absorber [29] to design of multi-layer optical thin films for color generation [30].

It would be of great interest to utilize a DRL-enabled photonic device design for the systematic optimization of the DRRs for the maximal DBP, which is the key unit to build an advanced on-chip optical reservoir, for all-optical packet header recognition.

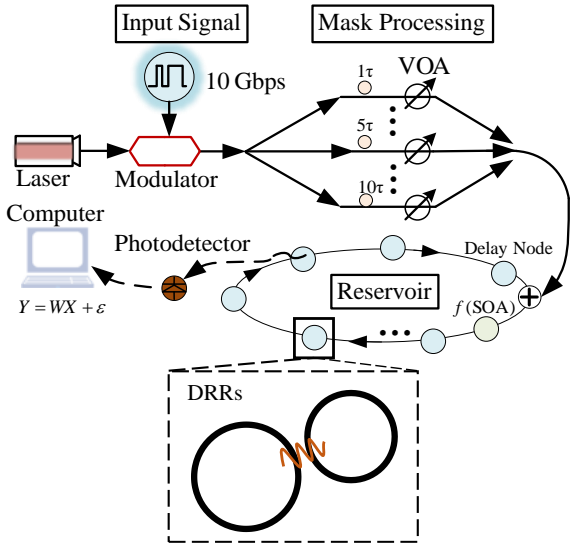

**Fig. 1.** Packet header recognition system with an all-optical reservoir. The reservoir consists of a nonlinear activation unit (green node) based on a SOA, a photodetector, and optical delay nodes (blue nodes), which are DRRs. VOA: variable optical attenuator; SOA: semiconductor optical amplifier.

In this paper, we propose an all-optical reservoir with DBP-optimized DRRs as the nodes for packet header recognition, as shown in Fig. 1. The DBP of four types of DRRs (shown in Fig. 2) are systematically optimized using an improved DRL method. After training, the algorithm can find the global maximal DBP for the DRRs in a vast structural parameter space within only 5 minutes. The optimal DBP found through this algorithm is better than the results of most traditional algorithms. Intriguingly, we note that the optimized double rings in cascaded, parallel, and embedded configurations have the same maximal DBP value of 1395 ps·GHz, more than twice that of the 3×3 coupler-based DRRs. These results suggest the optimal DBP is a global maximum and also deepen our understanding of the DRRs. The all-optical reservoir built with the optimized cascaded DRRs enables a greatly reduced word-error rate (WER) to 5×10⁻⁴ and 9×10⁻⁴ for 3-bit and 6-bit packet header recognition tasks, respectively, which are one order of magnitude lower compared to previously reported results. Our DRR-based all-optical reservoir also provides an integrated approach for optical header recognition, which significantly improves the scalability of the system with reduced power-consumption.

## II. OPTIMIZATION OF THE DRRS AS NODES OF AN ALL-OPTICAL RESERVOIR

Here, we consider four types of DRRs and improve the asynchronous advantage actor-critic (A3C) algorithm to maximize their DBP and use them as the delay nodes in an all-optical reservoir. The optimized DBP for each type of the DRRs is then analyzed. We also compare the proposed improved A3C algorithm to others at the end of the section.

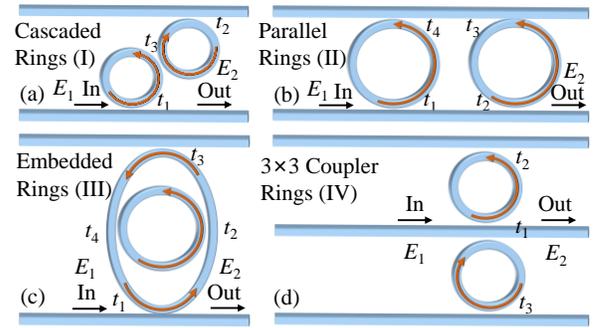

**Fig. 2.** Schematic of four types of DRRs. (a) cascaded rings (I), (b) parallel rings (II), (c) embedded rings (III), and (d) 3×3 coupler-based rings (IV).

### A. Categorization of the DRRs

In general, DRRs can be categorized into four types: (a) cascaded rings [26], (b) parallel rings [15], (c) embedded rings [34], and (d) 3×3 coupler-based rings [35], as shown in Fig. 2. Two microrings are coupled to each other or bus waveguides, with amplitude coupling coefficients labeled as $t_i$ ($i$ = 1 to 4). Note that the sizes and positions of the ring resonators and the coupling regions to the bus waveguides in each type of DRRs in Fig. 2 can be tailored. In this way, almost all of DRRs, although placed in various configurations, can be viewed as a variant of one of the four types above. A good example can be a DRR structure coupled to a cross-connect DRRs [36], which is a variant of the cascaded rings in Fig. 2(a) with a waveguide rotated.

Here, we consider that the device is comprised of thin-film lithium niobate waveguides. For a fair comparison, we fix waveguide loss at 0.1 dB/cm in all cases with an effective refractive index of 1.9. Since a large circumference in a single ring resonator results in a narrow linewidth and a large delay, with a fixed DBP [24], for the four types of the DRRs above, one should keep the sum of the circumferences of the two rings the same (here the sum is 880 $\mu$m), and thus two rings may have different resonance wavelengths. In this way, we form a comparison for all DRRs, in which an optical waveguide with a certain effective index (e.g., 1.9 here) is fabricated in some way with a certain loss (e.g., 0.1 dB/cm) and is used to build DRRs, with arbitrary coupling strengths and round trips of the two resonators, as long as the sum of their circumferences is kept fixed. The variables to be optimized in the DRRs are the coupling coefficients ($t_i$) and waveguide lengths between adjacent coupling regions.

The DBP is calculated from the transfer function ($T$) of a DRR system, using coupled mode theory [16], in which $T$ = $E_2/E_1$, where $E_1$ and $E_2$ are optical fields at the 'In' and 'Out' ports, respectively, as shown in Fig. 2.

### B. Improved A3C Algorithm

As a DRL algorithm, the A3C algorithm combines the actor-critic networks and applies deep neural networks into multiple threads for synchronous training. We improve the A3C algorithm to optimize DRRs, as shown in Fig. 3, by regularization of both entropy and reward functions. Thus, we call it regularization-assisted A3C algorithm (RA-A3C).

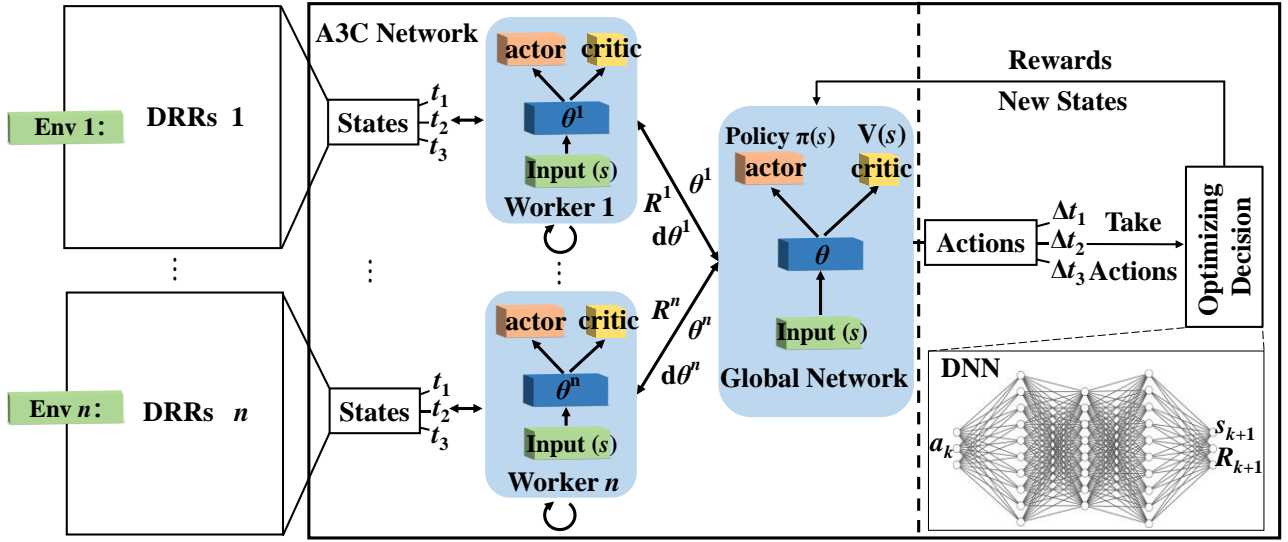

**Fig. 3.** The RA-A3C reinforcement learning algorithm. Coupled DRRs are the environment interacting with the A3C network. A3C Network consists of *n* workers and a global network. The global network summarizes the experience, and distributes neural network parameters and new states to each worker. The actor network (orange cube) and the critic network (yellow cube) correspond to the process of Policy $\pi(s)$ and V($s$), respectively. DNN: deep neural networks.

DRRs are treated as the environment of the RA-A3C algorithm. Here, we consider parallel optimization of DRRs in *n* different initial configurations, corresponding to *n* workers in the A3C network, in which the states ($s$) and actions ($a$) are the coupling coefficients and the changes of coupling coefficients, respectively. Global network and workers in the A3C network have the same architecture and parameters. The difference between them is that the global network only collects the experience passed by the workers to make decisions, and does not interact with real environments. We assume the number of workers $n = 12$. The reward of the RA-A3C algorithm is the sum of the DBP and the regularization term of states ($\lambda s$), as shown in (1):

$$R = DBP + \lambda s, \qquad (1)$$

where $\lambda$ is the regularization coefficient. When using the RA-A3C algorithm, training of the model can be completed only using initial values without a training set.

For the RA-A3C algorithm shown in Fig. 3, the actor is represented by the policy $\pi(a_k|s_k; \theta)$ and the critic is an estimate of the advantage function $V(a_k|s_k; \theta)$, where $\theta$ and $k$ are weight parameters and time step of the global network. In order to ensure correct iteration direction of the policy function and to prevent over-fitting of the neural network, entropy regularization and reward function regularization are implemented based on the A3C algorithm [38]. For a given policy $\pi(a_k|s_k; \theta)$, the Shannon entropy $H(\pi(a_k|s_k; \theta))$ is formulated as

$$H\left(\pi\left(a_k\,|s_k;\theta\right)\right) = -\sum_{i=1}^{M} P\left(a_i\,|s_i;\theta\right)\log P\left(a_i\,|s_k;\theta\right), \qquad (2)$$

where the logarithm is computed elementwise over the probabilities of the policy vector. $\pi(a_k|s_k; \theta) = [P(a_1|s_k; \theta), ..., P(a_M|s_k; \theta)]$ for an action set of *M* actions. Note that the probabilities ($P$) are parameterized by $\theta$ corresponding to the policy-specific weights in the global neural network.

Hence, the entropy $H(\pi(a_k|s_k; \theta))$ can be used as a tool to motivate the agent to steer clear of less non-deterministic policies, which is the entropy regularization [39]. Taking into consideration the entropy, we define the cost function as

$$f\left(\theta\right) = \log \pi\left(a_k\,|s_k;\theta\right) * V\left(a_k\,|s_k;\theta\right) * \sum R + \beta H, \qquad (3)$$

where $\beta$ is the weight of the entropy regularization. Additionally, regularization of the reward function is used when updating parameters of the global network,

$$d\theta = d\theta + \nabla_\theta f\left(\theta\right) + \eta \sum R, \qquad (4)$$

where $\eta$ is the reward function regularization hyperparameter, $\theta$ are weight parameters of the global network.

### C. Optimized DBP for the Four Types of DRRs

We use the RA-A3C algorithm to maximize the DBP of four types of DRRs and obtain the transmission and group delay spectra for the four types of DRRs in their optimized configurations, respectively, as shown in Fig. 4. We note that the cascaded rings and embedded rings show desirable "flat-top" transmission and delay spectra around the resonance wavelength.

The DBP of the cascaded rings reaches a maximum of 1395 ps·GHz with a delay of 27.5 ps and a bandwidth of 50.7 GHz, as shown in Fig. 4. Coupling coefficients for the optimized cascaded DRRs are $t_1 = 0.90$, $t_2 = 0.01$, and $t_3 = 0.35$.

Intriguingly, as shown in Table I, the maximal DBPs of the cascaded rings, parallel rings, and embedded rings reach the same value of 1395 ps·GHz, which is also an indicator that the global maximum is found. This value is about twice that of the 3×3 coupler-based rings (622 ps·GHz), suggesting this structure is inherently different from the other three types of the DRRs. In other words, the 3×3 coupler-based DRRs function more like a single resonator, in which its DBP is found to be 603 ps·GHz after optimization. It is also important to note that the first three types of the DRRs exhibits a DBP maximum more than twice that of a single resonator with a circumference equal to half of the circumference sum in the DRRs. It would be of interest to explore whether or not the maximal DBP can nonlinearly increase as the number of rings linearly increases, but this is beyond the scope of this paper.

TABLE I
THE MAXIMUM DBP AND OPTIMIZED COUPLING COEFFICIENTS OF FOUR TYPES OF DRRS

| Cases | $t_1$ | $t_2$ | $t_3$ | $t_4$ | Delay | DBP |
|---|---|---|---|---|---|---|
| cascaded DRRs (I) | 0.90 | 0.01 | 0.35 | - | 27.5 ps | 1395 ps·GHz |
| parallel DRRs (II) | 0.88 | 0.88 | 0.00 | 0.00 | 25.0 ps | 1395 ps·GHz |
| embedded DRRs (III) | 0.90 | 1.00 | 0.00 | 0.65 | 27.8 ps | 1395 ps·GHz |
| 3×3 coupler-based DRRs (IV) | 0.67 | 0.00 | 0.00 | - | 19.1 ps | 622 ps·GHz |

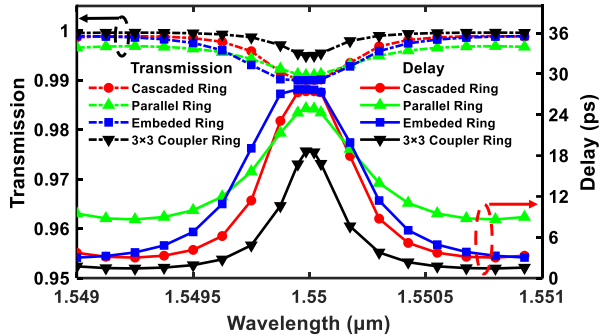

Fig. 4. The transmission and group delay spectra for the four types of DRRs in DBP-optimized configurations using the RA-A3C algorithm.

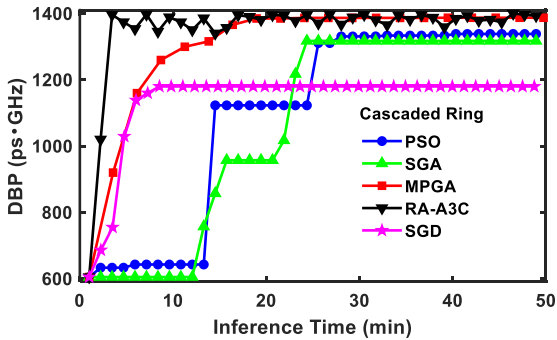

Fig. 5. For the cascaded rings, the optimized delay-bandwidth product (DBP) over the inference time using different algorithms. To ensure fast optical modelling and reinforcement learning procedure at the same time, the programs are performed on Python 3.7 on a Linux x64 server with a 20-core 2.40 GHz CPU processor, 128 GB of RAM, and a NVIDIA RTX 3070 GPU, and the model parameters are updated using the Adam optimizer [30].

### D. Algorithm Performance Comparison

To compare the efficiency of different algorithms, we take cascaded DRRs as an example and show the optimized DBP over inference time for various algorithms in Fig. 5. For a fair comparison, starting values of all the algorithms are set to the same random seed.

As shown in Fig. 5, the inference time of the proposed RA-A3C algorithm is less than 5 minutes, whereas most of the traditional optimization algorithms, including particle swarm optimization algorithm (PSO), simple genetic algorithm (SGA), multi-population genetic algorithm (MPGA), and stochastic gradient descent algorithm (SGD), take more than 20 minutes. For the DBP optimization problem of the DRRs, the inference time is a significant indicator. When the initial parameters are changed, traditional algorithms need to be re-run for DRRs in different initialization states, because they do not have the ability to "learn".

In contrast, knowledge learned from previous optimizations helps the RA-A3C algorithm to find the best solution in fewer steps, dramatically increasing the efficiency of the algorithm. In addition, the optimal DBP found with the RA-A3C algorithm (1395 ps·GHz) is better than the results using most traditional algorithms (e.g., SGD, 1186 ps·GHz), as in Fig. 5, which means that convergence error of the RA-A3C algorithm is much lower than those of the traditional algorithms. The inference time of the RA-A3C algorithm may be further decreased to the order of seconds by reducing the number of layers and neurons in the neural network or using the pruning strategy [40], [41].

### III. OPTICAL PACKET HEADER RECOGNITION WITH ALL-OPTICAL RESERVOIR

In this section, we use the optimized DRRs to build an all-optical reservoir for all-optical packet header recognition. Its bandwidth needs to be larger than the signal bandwidth. It is also desirable for the nodes in the reservoir to have a large group delay so that the footprint of the reservoir can be reduced significantly. From above, both the cascaded rings and embedded rings show a large DBP value and the desirable "flat-top" transmission and delay profiles around the resonant wavelength, which are favored as nodes for an all-optical reservoir for optical packet header recognition. Nevertheless, the optimized embedded rings require strong coupling in the ring-ring coupling regions, i.e., large $t_2$ and $t_4$ (refer to Table I), making the structure more difficult to fabricate than the cascaded rings. Hence, we consider the cascaded rings as the nodes of the reservoir in Fig. 1.

### A. Optical Packet Header Recognition System

The optical packet header recognition system shown in Fig. 1 works as follows: We use modulated laser signal to generate the incoming optical packet header signal. The header signal is firstly processed by mask, which is a function of enriching signal features by optical delays and variable optical attenuators. Then, the masked signal is sent to the optical reservoir consisting

## TABLE II
### 3-BIT PACKET HEADER RECOGNITION SYSTEM PARAMETER SETTINGS

| Parameters | Values | Parameters | Values | Parameters | Values |
|---|---|---|---|---|---|
| Signal rate | 10 Gbps | Delay of a loop ($\rho$) | 500 ps | Signal traveling time in the reservoir ($T$) | 3 ns |
| Number of bits | 3-bit | Node delay ($\tau_{\text{delay}}$) | 25 ps | Ridge coefficient ($\kappa$) | $8\times10^{-3}$ |
| Training data set | 6000 | Number of nodes ($N$) | 20 | WER | $5\times10^{-4}$ |
| Test data set | 2000 | Node bandwidth ($\Delta\omega$) | 50 GHz | - | - |

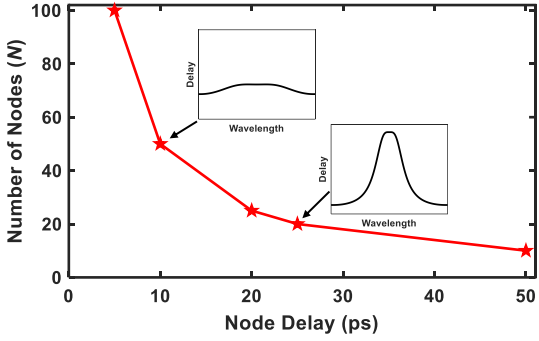

**Fig. 6.** Required number of nodes ($N$) for the node delay values ($\tau_{\text{delay}}$) is considered in our system, when the total time delay in the reservoir is a constant of 500 ps. Insets show the delay spectra for $\tau_{\text{delay}}$ = 10 ps and 25 ps, respectively.

of cascaded DRRs as nodes and a SOA as nonlinear activator. Lastly, when photodetector receives the optical packet header signals, it converts the complex amplitude into power, which also achieves the effect of a nonlinear activation unit. We receive the optical power signal from the photodetector and use ridge regression on the computer to realize data training process.

In our proposed experimental setup in Fig. 1. We set the signal rate to 10 Gbps, and consider a 10-channel mask. After mask processing, each bit in the optical signal is expanded to 10 different values with different delays. For data sets of 3-bit and 6-bit optical packet header signals, there are 8000 groups and 16000 groups of header signals, respectively. Parameter settings in our system related to the 3-bit header signal recognition task are summarized in Table II.

### B. Ridge Regression Algorithm and Indicators

After all neuron states in each node are collected, we adopt ridge regression algorithm to train the connection weights between the all-optical reservoir and the output.

Ridge regression algorithm is essentially the sum of linear regression and regularization terms, which is to solve the multi-collinearity problem that linear regression algorithm cannot deal with. Let $Y = Xw$ and $B$ be the actual and ideal output vectors. The purpose is to minimize the cost function $f$, and the cost function of ridge regression is formulated as

$$f = \left\| Xw - B \right\|^2 + \kappa \left\| w \right\|^2, \quad (5)$$

where $X$ is the input signal, $w$ is the weight of ridge regression, $\kappa$ is the ridge coefficient, and $\kappa = 8\times10^{-3}$. Taken the derivative of the cost function, $w$ can be calculated by

$$w = \left( X^T X + \kappa E \right)^{-1} X^T Y, \quad (6)$$

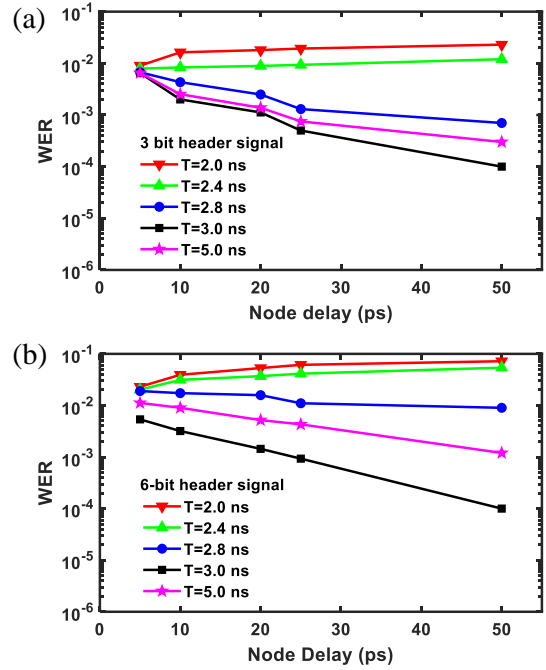

**Fig. 7.** WER in header recognition of (a) 3-bit and (b) 6-bit optical packet header signals. Signal traveling time in the reservoir ($T$) in the range of 2.0 ns to 5.0 ns is considered in both cases.

where $E$ is the identity matrix. Therefore, we can calculate the weights of ridge regression quickly.

To evaluate the accuracy and reliability of the recognition [5], we calculate WER as:

$$\text{WER} = \frac{N_{all} - N_{correct}}{N_{all}} \times 100\%, \quad (7)$$

where $N_{\text{all}}$ is the total number of the optical packet header signals, while $N_{\text{correct}}$ is the number of correctly recognized header signals.

### C. Reservoir Configuration Optimization and Results

Here, we optimize the reservoir configurations in terms of the node delay ($\tau_{\text{delay}}$) and the number of nodes ($N$) for a minimal WER. We vary $\tau_{\text{delay}}$ and $N$ and calculate the WER under a fixed total delay of a loop, i.e., $\rho = N \times \tau_{\text{delay}} = 500$ ps. We assume that signal traveling time in the reservoir ($T$) is an integer multiple of $\rho$. In our reservoir system, node delay $\tau_{\text{delay}}$ in the range of 5 ps to 50 ps is considered. In order to ensure that the total delay of a loop ($\rho$) stays the same, when the node delay increases, the number of nodes in the reservoir decreases accordingly. The number of nodes corresponding to each $\tau_{\text{delay}}$ value is shown in Fig. 6, and the delay spectra for $\tau_{\text{delay}}$ = 10 ps and 25 ps are

## TABLE III
## RECENTLY REPORTED RESULTS OF HEADER RECOGNITION

| Ref | Node type | Integrated or not | Bit | WER |
|---|---|---|---|---|
| [12] (Sim) | Virtual node | Yes | 3 | $1 \times 10^{-3}$ |
| [5] (Exp) | Fibers | Not | 6 | $1.3 \times 10^{-2}$ |
| [13] (Exp) | Spiral waveguide | Yes | 5 | $1 \times 10^{-2}$ |
| [6] (Exp) | Fibers | Not | 32 | $2 \times 10^{-3}$ |
| [11] (Exp) | FBG | Not | 2 | $1 \times 10^{-3}$ |
| **Our work (Sim)** | **DRRs** | **Yes** | **3** **6** | **$5 \times 10^{-4}$** **$9 \times 10^{-4}$** |

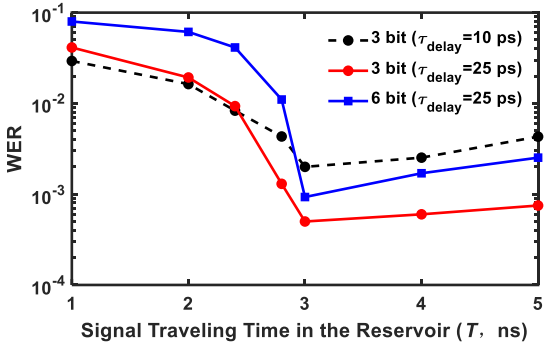

**Fig. 8.** WERs in 3-bit and 6-bit header recognition tasks as a function of signal traveling time in the reservoir ($T$) at $\tau_{\text{delay}} = 10$ ps and 25 ps for 3-bit signal and $\tau_{\text{delay}} = 25$ ps for 6-bit signal.

shown in the insets. This means that when the bandwidth is 50 GHz, the larger the node delay, the smaller the number of nodes in the reservoir.

The relationship between the WER and the node delay for 3-bit and 6-bit optical header signals is shown in Fig. 7. For both 3-bit and 6-bit cases, the WER increases with $\tau_{\text{delay}}$ when $T < 2.5$ ns. This feature is mainly attributed to the small number of samples, which leads to underfitting of the weight coefficients of the ridge regression. For $T > 2.5$ ns, WER decreases with $\tau_{\text{delay}}$, indicating an improved recognition accuracy.

Therefore, a larger node delay is favored for both 3-bit and 6-bit optical packet header recognition. Another advantage of using a larger node delay is that the system requires a smaller number of nodes. As a result, fewer weights are used for ridge regression. Fewer nodes also lead to a reduced footprint for the reservoir chip and a better system scalability.

For different traveling time of the optical packet header signal in the reservoir, we note that the WERs for both 3-bit and 6-bit header recognition first decrease and then slightly increase with $T$. We show this behavior of WERs in Fig. 8, for 3-bit signal with $\tau_{\text{delay}} = 10$ ps and 25 ps and 6-bit signal with $\tau_{\text{delay}} = 25$ ps as a function of $T$. The slight increase of WER for longer $T$ is mainly due to the optical loss in the reservoir. The number of weights for ridge regression is determined by physical quantities such as signal propagation time ($T$) in the reservoir. Hence, under-fitting or over-fitting will occur when $T$ is too small or too large. In our reservoir system, signal traveling time in the reservoir ($T$) in the range of 1 ns to 5 ns is considered. We would like to emphasize that a larger node delay is always favored for the optimal signal traveling time (e.g., $T = 3$ ns for 3-bit header signal), as indicated by the curves for $\tau_{\text{delay}} = 10$ ps and 25 ps cases in 3-bit header signal recognition.

Comparing WERs in 3-bit (red line) and 6-bit (blue line) head recognition tasks using the same node delay ($\tau_{\text{delay}} = 25$ ps) in Fig. 8, we notice that, for all the signal traveling time values in the reservoir considered here, the 3-bit signal recognition accuracy is always better than 6-bit. The reason is that, when the reservoir structure is the same, there are $2^3$ and $2^6$ different cases for the 3-bit and 6-bit header signals. Therefore, it is much more difficult for the reservoir to recognize the 6-bit signal than 3-bit. At $T = 3$ ns and $\tau_{\text{delay}} = 25$ ps, WERs of 3-bit and 6-bit optical packet header signals reach the minima, i.e., $5 \times 10^{-4}$ and $9 \times 10^{-4}$, respectively, 1~2 order of magnitude better than the previous results [5], [11], [13].

In Table III, our results are compared with previous works in terms of the node type, integration or not, signal rate, modulation format, and recognition accuracy. We note that many optical packet header recognition systems are either virtual systems or not integrated. Compared with the reported performance of the optical packet head recognition tasks, our system uses the DBP-maximized DRRs as the reservoir node has a better recognition accuracy among all the configurations. In addition, the experimental set-up containing the all-optical reservoir chip proposed in this work is relatively easy to realize, showing great promise for practical implementation.

## IV. CONCLUSION

In summary, we demonstrate 3-bit and 6-bit optical packet header recognition with an all-optical reservoir using DRRs as nodes. We utilize a RA-A3C reinforcement learning algorithm to maximize the DBP of the DRRs, to achieve the best performance of the optical reservoir. The optimized cascaded rings, parallel rings, and embedded rings can achieve the same DBP value of 1395 ps·GHz, which is larger than the optimized results using other algorithms. The all-optical reservoir for optical packet header recognition is formed with cascaded DRRs and trained via ridge regression. At a signal travel time $T = 3$ ns and a node delay $\tau_{\text{delay}} = 25$ ps, the WERs achieve optimized values of $5 \times 10^{-4}$ and $9 \times 10^{-4}$ for the 3-bit and 6-bit optical packet header signals, respectively, which are one order of magnitude smaller compared to those from previous works.

Our work reveals that, although placed in different configurations, various DRRs reach the same maximum DBP after optimization with the DRL algorithm, pointing to a global maximum for DRRs. Due to the key role of ring resonator devices in integrated photonics, exploring their full parameter space via AI algorithms would pave the way to miscellaneous applications apart from optical packet header recognition. For example, one may aim at optimizing the transmission slope for intensity-based optical sensors [42], [43] or maximizing the optical power enhancement in the cavity for nonlinear optical devices [44], to name a few.